\documentclass[12pt,english]{article}
\usepackage[T1]{fontenc}
\usepackage[latin1]{inputenc}
\usepackage{amsmath}
\usepackage{amssymb}
\usepackage{textcomp}
\usepackage[draft]{fixme}
\usepackage{natbib}
\usepackage{color}
\usepackage{multirow}

\newcommand{\bfbeta}{\mbox{\boldmath{$\beta$}}}
\newcommand{\bfepsilon}{\mbox{\boldmath{$\epsilon$}}}
\newcommand{\bftau}{\mbox{\boldmath{$\tau$}}}
\newcommand{\bfdelta}{\mbox{\boldmath{$\delta$}}}

\newcommand{\bfPsi}{\mbox{\boldmath{$\Psi$}}}
\newcommand{\bfLambda}{\mbox{\boldmath{$\Lambda$}}}
\newcommand{\bfsigma}{\mbox{\boldmath{$\sigma$}}}
\newcommand{\bfSigma}{\mbox{\boldmath{$\Sigma$}}}

\newcommand{\blind}{0}
  
\makeatletter

\usepackage{verbatim}

\usepackage{babel}
\makeatother
\begin{document}

\def\spacingset#1{\renewcommand{\baselinestretch}%
{#1}\small\normalsize} \spacingset{1}

\if0\blind
{
\title{Pure Error REML for Analyzing Data from Multi-Stratum Designs}

\author{Steven G.\ Gilmour\\
King's College London, UK
\and
Peter Goos\\
KU Leuven, Belgium\\
Universiteit Antwerpen, Belgium\\
\and
Heiko Gro{\ss}mann\\
Otto-von-Guericke-Universit\"{a}t Magdeburg, Germany}


\maketitle

} \fi

\if1\blind
{
  \bigskip
  \bigskip
  \bigskip
  \begin{center}
    {\LARGE\bf Pure Error REML for Analyzing Data from Multi-Stratum Designs}
\end{center}
  \medskip
} \fi
\bigskip

\begin{abstract}
Since the dawn of response surface methodology, it has been recommended that designs include replicate points, so that pure error estimates of variance can be obtained and used to provide unbiased estimated standard errors of the effects of factors. In designs with more than one stratum, such as split-plot and split-split-plot designs, it is less obvious how pure error estimates of the variance components should be obtained, and no pure error estimates are given by the popular residual maximum likelihood (REML) method of estimation. We propose a method of pure error REML estimation of the variance components, using the full treatment model, obtained by treating each combination of factor levels as a discrete treatment. Our method is easy to implement using standard software and improved estimated standard errors of the fixed effects estimates can be obtained by applying the Kenward-Roger correction based on the pure error REML estimates. We illustrate the new method using several data sets and compare the performance of pure error REML with the standard REML method. The results are comparable when the assumed response surface model is correct, but the new method is considerably more robust in the case of model misspecification.
\end{abstract}

\noindent \textbf{Keywords:} Full treatment model; Kenward-Roger correction; linear mixed model; replicates; response surface; split-plot design; split-split-plot design.

\newpage
\spacingset{1.45} 

\section{Introduction}\label{sec:intro}

It is increasingly recognized that many industrial and laboratory-based experiments are, or should be, run using split-plot and other multi-stratum structures, especially when some factors have levels which are harder to set than other factors. In an experiment, several \emph{treatments}, made up of combinations of levels of treatment factors, are applied under the control of the experimenter. The smallest unit to which a treatment can be applied is known as an \emph{experimental unit}, while the unit on which a response is observed is known as an \emph{observational unit}. In the simplest multi-stratum case, we define one or more easy-to-set factors whose levels can be reset for each experimental unit (often called a subplot or run), and one or more hard-to-set factors whose levels can only be reset for groups of experimental units (often called whole plots). A design is then chosen so that the hard-to-set factors' level combinations are randomized to whole plots, while the easy-to-set factors' level combinations are randomized to runs within whole plots. Such a design is called a split-plot design, and the whole-plot level and the sub-plot level of the split-plot design's structure are the two so-called strata of the design. 

The same principle is easily extended to more than two strata, which can be nested or crossed in any combination, in so-called multi-stratum designs. Provided they are continuous, the response data from such experiments are appropriately analyzed using linear or nonlinear mixed models, which include random effects for each stratum in the design and fixed effects of the treatments on the response. Often, the treatments' effects are modeled through a polynomial response surface model. The most common estimation procedure for such a response surface model uses residual maximum likelihood (REML), which maximizes the likelihood of a projection of the responses onto a subspace orthogonal to the assumed model, to estimate the variance components, and empirical generalized least squares (GLS), with the REML variance component estimates plugged in, to estimate the fixed parameters. This REML/GLS procedure is available in several statistical computing packages, gives the same results as analysis of variance in orthogonal multi-stratum designs and has good asymptotic properties. This analysis was recommended by \cite{lemy}, the first paper to systematically consider industrial split-plot response surface experiments.

However, it can often be observed in split-plot experiments that the whole-plot or block variance component is estimated to be zero. \cite{golava} showed that this is quite likely to happen even when the true value of the variance component is far from zero. \cite{gigo} discussed this problem and suggested a Bayesian analysis, which uses prior information on the variance components, as a reasonable alternative. This works well, but requires specialist software and careful specification of prior information. This prior information must be substantial to make up for the lack of information on the whole-plot or block variance component in the data.

Rather than resorting to a Bayesian analysis, which requires specifying a prior distribution, experimenters might prefer a robust analysis, based on as few assumptions as possible. In particular, an analysis which is robust to lack of fit in the assumed response surface model is desirable. An important step in this direction was made in the papers by \cite{vikomo} and \cite{viko}, which recommended a simple analysis based on pure error estimation of each variance component obtained from replicate points, which possesses this robustness property. \cite{viko} recommended that all inference be done using these pure error estimates. The method they recommend, however, is only applicable to particular types of designs and only uses replicate points within whole plots and completely replicated whole plots to obtain pure error estimates. \cite{gitr} showed, in the context of blocked response surface designs, that this is a narrower definition of pure error than is used in completely randomized designs. The definition in completely randomized designs requires only the use of the full treatment model. Since split-plot designs are also incomplete block designs, with some main effects completely confounded with blocks, the arguments of \cite{gitr} apply to split-plot designs and more general multi-stratum designs as well. Hence, the methods we present in the current paper are applicable to blocked response surface designs, split-plot designs and any multi-stratum designs.

The purpose of this paper is to explore in generality the use of pure error estimates of variance components. We do this by using REML in combination with the full treatment model (instead of the polynomial response surface model). This approach, which we term pure error REML, maximizes the likelihood of a projection of the responses onto a subspace orthogonal to the full treatment model, to produce pure error estimates of the variance components. These estimates do not depend on the assumed response surface model form. This allows many designs which do not have explicit replication within whole plots, or repeated whole plots, to be analyzed using pure error estimates. Moreover, even when such replication is available, pure error REML makes more use of the information in the data than the procedure of \cite{viko}. More specifically, pure error REML also exploits other types of replicated treatments to obtain more precise variance component estimates. After obtaining the pure error REML estimates of the variance components, we use these estimates in empirical GLS to obtain a robust analysis which can be recommended for analyzing data from any multi-stratum design. 

The models discussed and the notation used in this paper are clarified in Section \ref{sec:models}. A modified REML/GLS method is introduced in Section \ref{sec:estimation} and applied to some data sets in Section \ref{sec:examples}. The properties of the new method are studied in Section \ref{sec:simulation}, both when the assumed polynomial response surface model is correct and when it is not. We finish with a discussion of some practical points in Section \ref{sec:discussion}.

\section{Models and Notation}\label{sec:models}

We refer to any factor which divides the runs into groups as a ``blocking factor'', so that blocks, whole plots and subplots in a split-split-plot design are all referred to as blocking factors. In any experimental design, blocking factors arise as restrictions to the randomization, so that some sets of treatments must appear in runs that are in the same block. Unless every block consists of the same set of treatments, some information for comparing treatments is confounded with block effects, even when all parameters of the assumed response surface model are estimated orthogonally to block effects. If there are nested blocking factors, e.g.\ runs within blocks within superblocks, information concerning the treatment effects can be confounded with both the block effects and the superblock effects. By assuming that the block effects are random, the information available in the block totals can be recovered and combined with the usual within-block information to obtain more precise estimates of the treatment effects. Each level of blocking, e.g.\ blocks, superblocks, etc., leads to a \emph{stratum} in the analysis. Split-plot designs are block designs with at least one main effect completely confounded with blocks (usually called whole plots) and split-split-plot designs are nested block designs with at least one main effect confounded with superblocks (usually called whole plots) and at least one main effect completely confounded with blocks within superblocks (usually called subplots). More generally, multi-stratum designs involve treatments that are defined by combinations of the levels of several treatment factors applied in different strata. Therefore, some treatment factors have main effects that are confounded with the effects of some blocking factors. 

The model derived from the randomization, following \cite{hike}, is
\begin{equation}
\mathbf{Y} = \mathbf{X}_t\bftau +\sum_{j=1}^s\mathbf{Z}_j\bfdelta_j +\bfepsilon,
\label{eq:full}
\end{equation}
where $\mathbf{Y}$ is a random variable of which the response vector $\mathbf{y}$ is assumed to be a realization, $\mathbf{X}_t$ is
the $n\times t$ full treatment design matrix, having $(i,r)$th element equal to $1$ if treatment $r$ appears in run $i$ and $0$ otherwise, $n$ is the number of runs, $t$ is the number of treatments, $\bftau$ is the corresponding
vector of treatment means, $s$ is the number of blocking factors (which implies there are $s+1$ strata), $\bfdelta_j$ is the vector of random block effects
corresponding to the $j$th stratum, $\mathbf{Z}_j$ is the design matrix for these random effects and $\bfepsilon$ is the vector of random experimental unit
errors. We further assume that $\bfdelta_j\sim N\left(\mathbf{0}, \sigma_j^2 \mathbf{I}\right)$ with dimension $n_j$, the number of units in stratum $j$, $\bfepsilon\sim N\left(\mathbf{0}, \sigma^2\mathbf{I}\right)$ with dimension $n$ and $\bfdelta_j$, $j=1,\ldots,s$, and $\bfepsilon$ are mutually independent. We refer to model (\ref{eq:full}) as the full treatment model. The key feature of the full treatment model is that every factor level combination is viewed as one level of a single categorical factor. That level is called a \emph{treatment}.

In a typical response surface experiment, we want to interpret the treatment effects in more detail, for example by assuming that
\begin{equation}
\mathbf{X}_t\bftau = \mathbf{X}\bfbeta, \label{eq:polynomial}
\end{equation}
where $\mathbf{X}$ is the $n\times p$
design matrix for a second-order polynomial response surface model, $p = 1+2q+q(q-1)/2$ is the number of parameters in the response surface model, $q$ is the number of treatment factors and $\bfbeta$ is the vector of parameters of this model. Obviously, adopting the polynomial model is a much stronger assumption than
assuming the full treatment model in Equation~(\ref{eq:full}), which allows any pattern of treatment effects. Therefore, we would like our analysis to
depend on the assumed polynomial model as little as possible. This is consistent
with the presentation of \cite{bodr2}, who emphasize that
the popular polynomials are purely empirical graduating
functions. Throughout this paper, we use the second-order polynomial response surface model, but the ideas we present apply equally to any other linear regression model.

\section{Estimation}\label{sec:estimation}

In response surface studies, the main interest is usually in estimating the fixed effects $\bfbeta$ in the polynomial response surface model in Equation~(\ref{eq:polynomial}). If the ratios of variance components were known, this would be done optimally using generalized least squares (GLS), $\hat{\bfbeta}= (\mathbf{X}^\prime\bfSigma^{-1}\mathbf{X})^{-1} \mathbf{X}^\prime\bfSigma^{-1}\mathbf{Y},$
where
\begin{equation}
\bfSigma= \sigma^2 \left(\sum_{j=1}^s\gamma_j\mathbf{Z}_j\mathbf{Z}_j^\prime+ \mathbf{I}\right) \label{eq:Sigma}
\end{equation}
and $\gamma_j= \sigma_j^2/ \sigma^2$. The variance-covariance matrix of the GLS estimator is 
\[
\mathbf{V}(\hat{\bfbeta})= \bfPsi= (\mathbf{X}^\prime\bfSigma^{-1}\mathbf{X})^{-1}.
\]
In practice, of course, the ratios $\gamma_j$ of the variance components are not known, but have to be estimated. The particular method used to estimate the variance components then has an impact on the estimates obtained for the fixed effects. Plugging the variance components into Equation~(\ref{eq:Sigma}) and inserting the resulting matrix in the expression for $\hat{\bfbeta}$ produces the empirical GLS estimates of the fixed effects.

\subsection{Estimating Variance Components}\label{secvarcompest}

Following \cite{lemy}, it has become accepted that the variance components in multi-stratum response surface designs should be estimated using REML and the fixed effects should be estimated by empirical GLS. We refer to this approach as the REML/GLS approach. The main idea behind the pure error REML approach we introduce in this paper is to avoid the excessive reliance on the polynomial response surface model in Equation~(\ref{eq:polynomial}) made by the usual
REML/GLS analysis, because that model may be incorrectly specified. We do this by applying REML to the full treatment model in Equation~(\ref{eq:full}) and using
the resulting variance component estimates in the subsequent GLS analysis based on the polynomial model in Equation~(\ref{eq:polynomial}). One advantage of
this approach is that it produces pure error estimates of the
variance components, as recommended by \cite{viko}. We plug the REML variance component estimates from the full treatment model in Equation~(\ref{eq:full}) into the GLS estimator of the polynomial response surface model in Equation~(\ref{eq:polynomial}) and also use these variance component estimates to calculate the estimated standard errors of the fixed effects.

The argument in favor of the proposed analysis is analogous to that in completely randomized designs as to whether estimated standard errors should be based on the pure error estimate of the error variance or on the estimate for the error variance obtained from the polynomial regression model. According to \cite{bohu}, it is clearly advisable to use the pure error estimates when they are available. The common practice of checking for lack of fit and using the variance component estimates from the polynomial response surface model if no lack of fit is found can be thought of as an approximation to this. In that approach, the variance component estimates from the polynomial response surface model are only used if they are similar to the pure error estimates. The reason to use the variance component estimates from the polynomial response surface model, despite the fact that the pure error estimates are similar, is to increase the degrees of freedom for estimating error and obtain apparently more precise estimates and more powerful tests. Of course, this increase in precision is spurious due to the model selection involved. 

In any case, in a completely randomized response surface experiment, few would recommend ignoring the separation of pure error from lack of fit \citep{bodr2, mymoac}. However, this is precisely what is done in multi-stratum response surface designs when the empirical GLS estimates are obtained using the variance component estimates produced by applying REML to the polynomial regression model. This is even worse than in a completely randomized design, since, in multi-stratum designs, not only the estimated standard errors depend on the variance component estimates, but also the estimates of the fixed effects.

The REML method we recommend for estimating the variance components in the full treatment model in Equation~(\ref{eq:full}) involves maximizing the likelihood of the residuals, after removing the full treatment model's fixed effects. Specifically, the likelihood of $\mathbf{K}^\prime\mathbf{Y}$ is maximized, where $\mathbf{K}^\prime\mathbf{X}_t= \mathbf{0}$ and $rank(\mathbf{K})= n-t$, where $t$ is the number of treatments - see, for example, \cite{mcsene}. Unlike maximum likelihood estimators, REML estimators of the variance components are unbiased. Moreover, they are identical to the analysis of variance estimators in the case of orthogonal multi-stratum designs. 

Following \cite{lemy}, REML is currently usually applied to the polynomial regression model in Equation~(\ref{eq:polynomial}) to obtain estimators of the variance components, i.e.\ with $\mathbf{X}_t$ replaced by $\mathbf{X}$ and $\mathbf{K}$ having rank $n-p$ instead of $n-t$. In that case, the REML estimates are based on more degrees of freedom, but they depend on the assumed polynomial regression model, which is not necessarily a good approximation of the correct treatment model.

\subsection{Estimating Fixed Effects}

Estimating the treatment factors' effects is usually done using the empirical GLS estimator
\begin{equation}\label{eq:estpoly}
\hat{\bfbeta}= (\mathbf{X}^\prime\hat{\bfSigma}^{-1}\mathbf{X})^{-1} \mathbf{X}^\prime\hat{\bfSigma}^{-1}\mathbf{Y},
\end{equation}
where
\[
\hat{\bfSigma}= \hat{\sigma}^2 \left(\sum_{j=1}^s\hat{\gamma}_j\mathbf{Z}_j\mathbf{Z}_j^\prime+ \mathbf{I}\right),
\]
$\hat{\gamma}_j= \hat{\sigma}_j^2/ \hat{\sigma}^2$ and $\hat{\sigma}_j^2$ and $\hat{\sigma}^2$ are obtained from REML applied to the polynomial response surface model instead of the full treatment model. The variance matrix of these estimators is usually estimated by
\begin{equation}\label{eq:estvar}
\hat{\mathbf{V}}(\hat{\bfbeta})= \hat{\bfPsi}= (\mathbf{X}^\prime\hat{\bfSigma}^{-1}\mathbf{X})^{-1}.
\end{equation}

The focus of this paper is on the plug-in estimators of the variance components which are used in the empirical GLS estimator in Equation~(\ref{eq:estpoly}). The most crucial property these estimators must have is that the estimated variance components are as close to the true variance components as possible. There is no theoretical reason why the plug-in estimators should be obtained from the polynomial response surface model in Equation~(\ref{eq:polynomial}), even though this is the model whose fixed effects we are estimating. We suggest that the unbiased estimators of the variance components obtained from the full treatment model in Equation~(\ref{eq:full}) should be used instead. We will refer to this as the pure error REML/GLS, or PE-REML/GLS, method; the standard REML method, based on the polynomial response surface treatment model, will be referred to as RS-REML/GLS.

The problem with using the usual estimators of the variance components in Equation~(\ref{eq:estpoly}) is that they are used both to obtain the fixed effects estimates and to assess the quality of these estimates using the variance matrix in Equation~(\ref{eq:estvar}). Consider, for example, a situation in which the polynomial model is inadequate because some higher-order terms are missing. Then, the estimated variance components from this model are likely to be overestimated, perhaps considerably so. This leads to overestimated standard errors of the fixed effect estimates, which, in turn, might lead to few effects seeming to be significantly different from zero. We might then make the decision to reduce the order of the model, rather than to increase it. If, on the other hand, the variance components are estimated from the full model, they will be unbiased irrespective of whether the assumed polynomial model is correct or not. The resulting analysis should then correctly show the inadequacy of the assumed polynomial response surface model, e.g.\ by using the lack of fit test proposed by \cite{goosgilmour16}. Of course, since the unbiased estimators are based on less information (i.e.\ fewer degrees of freedom) than the biased estimators, things are not quite so simple. However, this line of reasoning does at least indicate that, in the case of major model inadequacies, the unbiased estimators should be better.

\subsection{Estimating Standard Errors of Fixed Effects}\label{sec:SEs}

The estimated variances and standard errors of the fixed effects' estimates obtained using Equation~(\ref{eq:estvar}) are known to be negatively biased even if the fixed-effects model is correct. Therefore, we use the Kenward-Roger correction, \citep{kero}. By a direct application of results given by \cite{mcsene} (p.\ 165-169), the Kenward-Roger correction is trivially adapted to the PE-REML method, i.e.\ it can be applied regardless of which estimator of the variance components is used for estimating the fixed effects.

In simple orthogonal block structures (\cite{nelder}; see also \cite{gitr3}), with the polynomial response surface model in Equation~(\ref{eq:polynomial}) for the fixed effects and applying results analogous to those of \cite{goosgilmour16}, the approximate variance matrix for fixed effects estimates, with the Kenward-Roger correction, is
\begin{equation}\label{eq:SigmaInv}
\widehat{\hat{\mathbf{V}}(\hat{\bfbeta})}
= \hat{\bfPsi}+ 2\hat{\bfLambda},
\end{equation}
where $\hat{\bfPsi}$ is from Equation~(\ref{eq:estvar}), $\hat{\bfLambda}$ is obtained by plugging the appropriate REML estimates $\hat{\bfsigma}$ of the variance components into
\[
\hat{\bfLambda} = \hat{\bfPsi} \left\{\sum_{i=1}^{s+1}\sum_{j=1}^{s+1}\hat{u}_{ij} \left(\hat{\mathbf{Q}}_{ij}-\hat{\mathbf{P}}_i\hat{\bfPsi}\hat{\mathbf{P}}_j\right)\right\} \hat{\bfPsi},
\]
$\hat{u}_{ij}$ is the $(i,j)$th element of the estimated variance matrix of $\hat{\bfsigma}$, either the RS-REML or PE-REML estimator of $\bfsigma= [\sigma_1^2~\cdots~\sigma_{s+1}^2]^\prime$, $\sigma_{s+1}^2 = \sigma^2$,
\[
\mathbf{P}_i= \mathbf{X}_g^\prime \frac{\partial \bfSigma^{-1}}{\partial \sigma_i^2} \mathbf{X}_g,
\]
\[
\mathbf{Q}_{ij}= \mathbf{X}_g^\prime \frac{\partial \bfSigma^{-1}}{\partial \sigma_i^2} \bfSigma \frac{\partial \bfSigma^{-1}}{\partial \sigma_j^2} \mathbf{X}_g,
\]
$\mathbf{X}_g$ is the design matrix corresponding to some fixed-effects model (either $\mathbf{X}_t$ or $\mathbf{X}$ corresponding to the full treatment model or the polynomial response surface model, respectively) and $\bfSigma$ is defined in (\ref{eq:Sigma}). The $\hat{u}_{ij}$ are obtained by plugging the estimated variance components into the inverse of the diagonal block corresponding to the variance components of the Fisher information matrix based on the RS-REML or PE-REML likelihood function respectively - see \citet[p.\ 177-178]{mcsene}.  The Kenward-Roger correction affects only those standard errors which are influenced by the plug-in estimators of $\sigma_1^2, \ldots, \sigma_s^2$ and not those which depend only on $\sigma^2$.

In the particular case of a split-plot design with $k$ subplots within each whole plot, so that  $s=1$ and $\mathbf{Z}_1$ in (\ref{eq:full}) is denoted by $\mathbf{Z}$, for both RS-REML and PE-REML we have $\hat{u}_{11}= 2tr(\mathbf{CC})/c,$ $\hat{u}_{22}= 2tr(\mathbf{Z}^\prime\mathbf{CZZ}^\prime\mathbf{CZ})/c$
and $\hat{u}_{12} = \hat{u}_{21} =  -2tr(\mathbf{Z}^\prime\mathbf{CC}\mathbf{Z})/c,$
where $c= tr(\mathbf{CC})tr(\mathbf{Z}^\prime\mathbf{CZZ}^\prime\mathbf{CZ})- \left\{tr(\mathbf{Z}^\prime\mathbf{CC}\mathbf{Z})\right\}^2$
and
\[
\mathbf{C}= \hat{\bfSigma}^{-1}- \hat{\bfSigma}^{-1}\mathbf{X}_g \left(\mathbf{X}_g^\prime\hat{\bfSigma}^{-1}\mathbf{X}_g\right)^{-1} \mathbf{X}_g^\prime\hat{\bfSigma}^{-1}.
\]
We also have
\[
\hat{\mathbf{P}}_1 = -\frac{1}{\left(\hat{\sigma}^2+k\hat{\sigma}_1^2\right)^2} \mathbf{X}_g^\prime  \mathbf{ZZ}^\prime \mathbf{X}_g,
\]
\[
\hat{\mathbf{P}}_2 = \frac{1}{\hat{\sigma}^4} \mathbf{X}_g^\prime \left\{\frac{\hat{\sigma}_1^2\left(2\hat{\sigma}^2+k\hat{\sigma}_1^2\right)}{\left(\hat{\sigma}^2+k\hat{\sigma}_1^2\right)^2} \mathbf{ZZ}^\prime- \mathbf{I}\right\} \mathbf{X}_g,
\]
\[
\hat{\mathbf{Q}}_{11} = \frac{1}{\left(\hat{\sigma}^2+k\hat{\sigma}_1^2\right)^4} \mathbf{X}_g^\prime  \mathbf{ZZ}^\prime \hat{\bfSigma} \mathbf{ZZ}^\prime \mathbf{X}_g,
\]
\[
\hat{\mathbf{Q}}_{22} = \frac{1}{\hat{\sigma}^8} \mathbf{X}_g^\prime \left\{\frac{\hat{\sigma}_1^2\left(2\hat{\sigma}^2+k\hat{\sigma}_1^2\right)}{\left(\hat{\sigma}^2+k\hat{\sigma}_1^2\right)^2} \mathbf{ZZ}^\prime- \mathbf{I}\right\} \hat{\bfSigma} \left\{\frac{\hat{\sigma}_1^2\left(2\hat{\sigma}^2+k\hat{\sigma}_1^2\right)}{\left(\hat{\sigma}^2+k\hat{\sigma}_1^2\right)^2} \mathbf{ZZ}^\prime- \mathbf{I}\right\} \mathbf{X}_g
\]
and
\[
\hat{\mathbf{Q}}_{12} = \hat{\mathbf{Q}}_{21} = -\frac{1}{\hat{\sigma}^4\left(\hat{\sigma}^2+k\hat{\sigma}_1^2\right)^2} \mathbf{X}_g^\prime  \mathbf{ZZ}^\prime \hat{\bfSigma} \left\{\frac{\hat{\sigma}_1^2\left(2\hat{\sigma}^2+k\hat{\sigma}_1^2\right)}{\left(\hat{\sigma}^2+k\hat{\sigma}_1^2\right)^2} \mathbf{ZZ}^\prime- \mathbf{I}\right\} \mathbf{X}_g.
\]

One final point to note is that, if the estimated variance component for the blocks, $\hat{\sigma}_1^2$, is zero (or negative if this is allowed), then the Kenward-Roger approximation fails. In this case, we simply let $\hat{\bfLambda}= \mathbf{0}$ and the adjusted estimate is then equal to the unadjusted estimate. This is also what statistical software packages do. All of the above generalizes to any multi-stratum design, including any nested block structure. The results for a split-split-plot structure are given in the appendix.

\section{Illustrations}\label{sec:examples}

We illustrate the PE-REML method and compare it with the standard RS-REML method using three data sets. These were chosen to show how the new method works in different kinds of designs. The first thing to note is that many multi-stratum designs, especially D-optimal designs, do not allow pure error estimation of variance components, so that PE-REML estimation is infeasible and standard RS-REML estimation is the only possibility to obtain variance component estimates. The designs in our examples include an equivalent estimation split-plot design which has ample degrees of freedom for pure error, a split-plot design constructed by hand which has limited pure error degrees of freedom and an I-optimal split-split-plot design, which illustrates how the method extends to general multi-stratum structures. All three examples use the second-order polynomial model as the main model of interest. The coefficients $\beta_r$, $\beta_{rr}$ and $\beta_{rs}$ correspond to the linear effect of factor $X_r$, the quadratic effect of $X_r$ and the interaction of $X_r$ and $X_s$ respectively.

\subsection{Strength of Ceramic Pipes}

\begin{table}
\caption{Fixed-effect estimates and estimated standard errors for the ceramic pipe data.} \label{tab:ceramic}
\begin{center}
\begin{tabular}{cr|cc}
  \hline
\multirow{2}{*}{Parameter} & \multirow{2}{*}{Estimate} & \multicolumn{2}{c}{Standard Error} \\
& & RS-REML & PE-REML \\
  \hline
  $\beta_1$ & 4.5579 & 0.4893 & 0.3027 \\
  $\beta_2$ & $-$6.5592 & 0.4893 & 0.3027 \\
  $\beta_3$ & $-$4.9733 & 0.0648 & 0.0721 \\
  $\beta_4$ & 4.0922 & 0.0648 & 0.0721 \\
  $\beta_{11}$ & 1.7381 & 0.8974 & 0.5551 \\
  $\beta_{22}$ & $-$0.5407 & 0.8974 & 0.5551 \\
  $\beta_{33}$ & $-$2.3864 & 0.6059 & 0.3958 \\
  $\beta_{44}$ & 2.5736 & 0.6059 & 0.3958 \\
  $\beta_{12}$ & 0.8431 & 0.5993 & 0.3707 \\
  $\beta_{13}$ & 1.4356 & 0.0688 & 0.0765 \\
  $\beta_{14}$ & $-$1.4794 & 0.0688 & 0.0765 \\
  $\beta_{23}$ & $-$1.0019 & 0.0688 & 0.0765 \\
  $\beta_{24}$ & 1.9856 & 0.0688 & 0.0765 \\
  $\beta_{34}$ & $-$1.0394 & 0.0688 & 0.0765 \\
   \hline
\end{tabular}
\end{center}
\end{table}

The design for the experiment on ceramic pipes reported by \cite{vikomo} has 12 whole plots, each with four runs, and three of the whole plots consisting of replicated center points. We name the coded factors in the experiment $X_1$ (zone 1 temperature), $X_2$ (zone 2 temperature), $X_3$ (amount of binder) and $X_4$ (grinding speed), where $X_1$ and $X_2$ are whole-plot factors. Because the design has the equivalent-estimation property (i.e.\ the OLS and GLS estimators are equivalent), both the RS-REML and PE-REML methods give the same estimates for the factor effects from empirical GLS. These estimates are shown in Table \ref{tab:ceramic} along with their estimated standard errors. In this case, the Kenward-Roger correction has no effect.

Using the full treatment model to estimate the variance components gives the same variance component estimates ($\hat{\sigma}_1^2 = 0.52626$ and $\hat{\sigma}^2 = 0.09355$), and, hence, the same estimated standard errors of the fixed effects as the sample variance method of \cite{vikomo} (although their Table 6 actually gives variances, wrongly labeled as standard errors). The estimated RS-REML standard errors from the polynomial regression model for the linear effects of the whole-plot factors $X_1$ and $X_2$ and their interaction and for quadratic effects are larger than those from the PE-REML method, while for the other effects they are smaller, but comparable. This is due to the fact that the RS-REML estimate of the variance component $\sigma^2_1$, $\hat{\sigma}_1^2 = 1.4176$, is larger than the PE-REML estimate, $\hat{\sigma}_1^2 = 0.52626$, while the RS-REML estimate of $\sigma^2$, $\hat{\sigma}^2 = 0.07563$, is smaller than the PE-REML estimate, $\hat{\sigma}^2 = 0.09355$.

The fact that the PE-REML method yields the same estimates as the sample variance approach of \cite{vikomo} shows that, whereas they contrasted their pure error standard error estimates with those obtained from REML, the point is not the method (REML or sample variances) used, but the treatment model used as a starting point for estimating variance components (the full treatment model rather than the RS model).

\subsection{Another split-plot example}\label{fakeex2}

\spacingset{1} 

\begin{table}
\caption{Split-plot design with 12 whole plots of five runs for estimating a second-order response surface model in two whole-plot factors and two subplot factors, along with simulated responses.} \label{fakedata2}
{\tiny \begin{center}
\begin{tabular}{ccrrrrr}
  \hline
\multicolumn{1}{c}{Whole plot} & \multicolumn{1}{c}{Treatment} & \multicolumn{1}{c}{$X_1$} & \multicolumn{1}{c}{$X_2$} & \multicolumn{1}{c}{$X_3$} & \multicolumn{1}{c}{$X_4$} & \multicolumn{1}{c}{$Y$} \\
  \hline
   1 &  1 & $-$1 &  $-$1 &  $-$1 &  $-$1 & 29.46 \\
   1 &  2 & $-$1 &  $-$1 &   1 &  $-$1 & 31.50 \\
   1 &  3 & $-$1 &  $-$1 &  $-$1 &   1 & 23.41 \\
   1 &  4 & $-$1 &  $-$1 &   1 &   1 & 19.12 \\
   1 &  5 & $-$1 &  $-$1 &   0 &   0 & 24.38 \\
	\hline
   2 &  6 &  1 &  $-$1 &  $-$1 &  $-$1 & 53.32 \\
   2 &  7 &  1 &  $-$1 &   1 &  $-$1 & 50.18 \\
   2 &  8 &  1 &  $-$1 &  $-$1 &   1 & 55.08 \\
   2 &  9 &  1 &  $-$1 &   1 &   1 & 47.97 \\
   2 & 10 &  1 &  $-$1 &   0 &   0 & 49.08 \\
	\hline
   3 & 11 & $-$1 &   1 &  $-$1 &  $-$1 & 37.10 \\
   3 & 12 & $-$1 &   1 &   1 &  $-$1 & 41.39 \\
   3 & 13 & $-$1 &   1 &  $-$1 &   1 & 43.22 \\
   3 & 14 & $-$1 &   1 &   1 &   1 & 38.18 \\
   3 & 15 & $-$1 &   1 &   0 &   0 & 38.85 \\
	\hline
   4 & 16 &  1 &   1 &  $-$1 &  $-$1 & 39.10 \\
   4 & 17 &  1 &   1 &   1 &  $-$1 & 44.05 \\
   4 & 18 &  1 &   1 &  $-$1 &   1 & 58.19 \\
   4 & 19 &  1 &   1 &   1 &   1 & 51.32 \\
   4 & 20 &  1 &   1 &   0 &   0 & 47.68 \\
	\hline
   5 & 21 & $-$1 &   0 &  $-$1 &   0 & 37.74 \\
   5 & 22 & $-$1 &   0 &   1 &   0 & 32.18 \\
   5 & 23 & $-$1 &   0 &   0 &  $-$1 & 37.27 \\
   5 & 24 & $-$1 &   0 &   0 &   1 & 34.25 \\
   5 & 25 & $-$1 &   0 &   0 &   0 & 36.18 \\
	\hline
   6 & 26 &  1 &   0 &  $-$1 &   0 & 49.91 \\
   6 & 27 &  1 &   0 &   1 &   0 & 50.84 \\
   6 & 28 &  1 &   0 &   0 &  $-$1 & 49.24 \\
   6 & 29 &  1 &   0 &   0 &   1 & 54.78 \\
   6 & 30 &  1 &   0 &   0 &   0 & 50.45 \\
	\hline
   7 & 31 &  0 &  $-$1 &  $-$1 &   0 & 40.63 \\
   7 & 32 &  0 &  $-$1 &   1 &   0 & 46.87 \\
   7 & 33 &  0 &  $-$1 &   0 &  $-$1 & 47.88 \\
   7 & 34 &  0 &  $-$1 &   0 &   1 & 42.95 \\
   7 & 35 &  0 &  $-$1 &   0 &   0 & 47.16 \\
	\hline
   8 & 36 &  0 &   1 &  $-$1 &   0 & 48.59 \\
   8 & 37 &  0 &   1 &   1 &   0 & 49.21 \\
   8 & 38 &  0 &   1 &   0 &  $-$1 & 48.14 \\
   8 & 39 &  0 &   1 &   0 &   1 & 53.42 \\
   8 & 40 &  0 &   1 &   0 &   0 & 49.59 \\
	\hline
   9 & 41 &  0 &   0 &  $-$1 &  $-$1 & 48.61 \\
   9 & 42 &  0 &   0 &   1 &  $-$1 & 51.91 \\
   9 & 43 &  0 &   0 &  $-$1 &   1 & 55.17 \\
   9 & 44 &  0 &   0 &   1 &   1 & 50.13 \\
   9 & 45 &  0 &   0 &   0 &   0 & 49.47 \\
	\hline
  10 & 46 &  0 &   0 &  $-$1 &   0 & 49.08 \\
  10 & 47 &  0 &   0 &   1 &   0 & 48.77 \\
  10 & 48 &  0 &   0 &   0 &  $-$1 & 51.00 \\
  10 & 49 &  0 &   0 &   0 &   1 & 49.52 \\
  10 & 45 &  0 &   0 &   0 &   0 & 49.72 \\
	\hline
  11 & 41 &  0 &   0 &  $-$1 &  $-$1 & 41.26 \\
  11 & 42 &  0 &   0 &   1 &  $-$1 & 43.83 \\
  11 & 43 &  0 &   0 &  $-$1 &   1 & 57.94 \\
  11 & 44 &  0 &   0 &   1 &   1 & 42.02 \\
  11 & 45 &  0 &   0 &   0 &   0 & 40.65 \\
	\hline
  12 & 46 &  0 &   0 &  $-$1 &   0 & 49.43 \\
  12 & 47 &  0 &   0 &   1 &   0 & 46.00 \\
  12 & 48 &  0 &   0 &   0 &  $-$1 & 54.96 \\
  12 & 49 &  0 &   0 &   0 &   1 & 55.10 \\
  12 & 45 &  0 &   0 &   0 &   0 & 44.45 \\
  \hline
\end{tabular}
\end{center}}
\end{table}

\spacingset{1.45} 

To illustrate some different points, we simulated data for another design involving four factors, two of which are applied in the whole-plot stratum, and 12 whole plots, each containing five runs. The design was constructed by hand, using standard treatment sets and the ideas of fractional partial confounding \citep{mead, mgm} to distribute the 49 treatments between whole plots - see Table \ref{fakedata2}. Unlike the design for the ceramic pipe experiment, this design has replicated treatments which appear only in different whole plots. All the replicated treatments, labelled 41-49, appear in whole plots 9-12. This example shows that it is possible to obtain pure error variance component estimates using the PE-REML method even from a design which does not have within-whole-plot replicates. This is in contrast to the sample variance method of \cite{viko} which does not allow pure error to be estimated from this design.


Responses were simulated using true values of the fixed effects in the second order model of $\beta_0 = 50$, $\beta_1 = 8$, $\beta_2 = 3$, $\beta_{11} = -7$, $\beta_{22} = -3$, $\beta_{44} = 1$, $\beta_{12} = -4$, $\beta_{14} = 2$, $\beta_{24} = 3$, $\beta_{34} = -2$ and all other parameters equal to zero, so that the assumptions of RS-REML are met in this case. The true values of the variance components were $\sigma_1^2=4$ and $\sigma^2 = 2$. Using PE-REML, the estimate of $\sigma_1^2$ is 5.3738 and the estimate of $\sigma^2$ is 10.552, while the RS-REML estimates are 3.1085 and 6.3957 respectively. In this case, $\sigma^2$ is not very well estimated by either method, even though there are adequate degrees of freedom for its estimation. 

\begin{table}
\caption{Fixed-effect estimates and standard errors for the simulated data from the split-plot design in Table~\ref{fakedata2}.} \label{tab:confoundest}
{\scriptsize \begin{center}
\begin{tabular}{crr|cccc}
  \hline
\multirow{2}{*}{Parameter} & \multicolumn{2}{c}{Estimate} & \multicolumn{4}{c}{Standard error} \\
& RS-REML & PE-REML & RS-REML & PE-REML & RS-REML-KR & PE-REML-KR \\
  \hline
  $\beta_1$ & 8.2320 & 8.2320 & 0.8551 & 1.1169 & 0.8551 & 1.1169 \\ 
  $\beta_2$ & 2.6347 & 2.6347 & 0.8551 & 1.1169 & 0.8551 & 1.1169 \\ 
  $\beta_3$ & $-$0.8825 & $-$0.8825 & 0.4215 & 0.5414 & 0.4215 & 0.5414 \\ 
  $\beta_4$ & 0.8769 & 0.8769 & 0.4215 & 0.5414 & 0.4215 & 0.5414 \\ 
  $\beta_{11}$ & $-$6.1579 & $-$6.1591 & 1.2865 & 1.6801 & 1.2867 & 1.6810 \\ 
  $\beta_{22}$ & $-$1.9979 & $-$1.9991 & 1.2865 & 1.6801 & 1.2867 & 1.6810 \\ 
  $\beta_{33}$ & $-$0.3846 & $-$0.3787 & 0.7137 & 0.9174 & 0.7245 & 0.9578 \\ 
  $\beta_{44}$ & 2.0538 & 2.0596 & 0.7137 & 0.9174 & 0.7245 & 0.9578 \\ 
  $\beta_{12}$ & $-$4.3080 & $-$4.3080 & 1.0473 & 1.3679 & 1.0473 & 1.3679 \\ 
  $\beta_{13}$ & $-$0.1340 & $-$0.1340 & 0.5655 & 0.7264 & 0.5655 & 0.7264 \\ 
  $\beta_{14}$ & 2.4995 & 2.4995 & 0.5655 & 0.7264 & 0.5655 & 0.7264 \\ 
  $\beta_{23}$ & 0.2105 & 0.2105 & 0.5655 & 0.7264 & 0.5655 & 0.7264 \\ 
  $\beta_{24}$ & 2.9180 & 2.9180 & 0.5655 & 0.7264 & 0.5655 & 0.7264 \\ 
  $\beta_{34}$ & $-$2.4283 & $-$2.4283 & 0.5162 & 0.6631 & 0.5162 & 0.6631 \\ 
\hline
\end{tabular}
\end{center}}
\end{table}

The fixed effects' estimates and their estimated standard errors, with and without the Kenward-Roger correction, are shown in Table~\ref{tab:confoundest}. The partially orthogonal structure of the design implies that the fixed effects estimates are the same for each method, except for the quadratic effects which differ slightly. As a result, the Kenward-Roger correction only affects the standard errors of the estimates of the quadratic effects. In this case, we find that the methods give very similar estimates, but slightly different standard errors. The standard errors obtained using the pure error estimates of the variance components are larger than those based on the polynomial regression model. This is because the pure error estimates of the variance components are larger than their RS-REML counterparts.

\subsection{A split-split-plot example}\label{fakeex3}

\spacingset{1} 

\begin{table}
\caption{Split-split-plot design with six whole plots, each containing two suplots of three runs for estimating a second-order response surface model in one whole-plot factor, one subplot factor and two subsubplot factors, along with simulated responses.} \label{sspdata}
\begin{center}
\begin{tabular}{cccrrrrr}
  \hline
\multicolumn{1}{c}{Whole plot} & \multicolumn{1}{c}{Subplot} & \multicolumn{1}{c}{Treatment} & \multicolumn{1}{c}{$X_1$} & \multicolumn{1}{c}{$X_2$} & \multicolumn{1}{c}{$X_3$} & \multicolumn{1}{c}{$X_4$} & \multicolumn{1}{c}{$Y$} \\
\hline
   1 & 1 & 1 & 0 & 0 & $-$1 & 1 & 50.77 \\ 
   1 & 1 & 2 & 0 & 0 & 0 & 0 & 49.08 \\ 
   1 & 1 & 2 & 0 & 0 & 0 & 0 & 50.21 \\ 
	\cline{2-8}
   1 & 2 & 2 & 0 & 0 & 0 & 0 & 47.28 \\ 
   1 & 2 & 3 & 0 & 0 & 1 & $-$1 & 48.64 \\ 
   1 & 2 & 2 & 0 & 0 & 0 & 0 & 49.18 \\ 
	\hline
   2 & 3 & 4 & 1 & $-$1 & 1 & 1 & 48.68 \\ 
   2 & 3 & 5 & 1 & $-$1 & $-$1 & $-$1 & 51.67 \\ 
   2 & 3 & 6 & 1 & $-$1 & 0 & 0 & 50.76 \\ 
	\cline{2-8}
	 2 & 4 & 7 & 1 & 0 & $-$1 & 0 & 52.02 \\ 
   2 & 4 & 8 & 1 & 0 & 1 & $-$1 & 52.36 \\ 
   2 & 4 & 9 & 1 & 0 & 0 & 1 & 54.37 \\ 
	\hline
   3 & 5 & 10 & $-$1 & $-$1 & 0 & 1 & 22.13 \\ 
   3 & 5 & 11 & $-$1 & $-$1 & 1 & $-$1 & 37.65 \\ 
   3 & 5 & 12 & $-$1 & $-$1 & $-$1 & 0 & 28.17 \\ 
	\cline{2-8}
	 3 & 6 & 13 & $-$1 & 0 & $-$1 & $-$1 & 37.88 \\ 
   3 & 6 & 14 & $-$1 & 0 & 1 & 1 & 35.33 \\ 
   3 & 6 & 15 & $-$1 & 0 & 0 & 0 & 37.38 \\ 
	\hline
   4 & 7 & 16 & 0 & $-$1 & 0 & $-$1 & 48.69 \\ 
   4 & 7 & 17 & 0 & $-$1 & $-$1 & 1 & 44.92 \\ 
   4 & 7 & 18 & 0 & $-$1 & 1 & 0 & 44.14 \\ 
  \cline{2-8}
	 4 & 8 & 19 & 0 & 1 & $-$1 & $-$1 & 46.72 \\ 
   4 & 8 & 20 & 0 & 1 & 1 & 0 & 50.06 \\ 
   4 & 8 & 21 & 0 & 1 & 0 & 1 & 54.69 \\ 
	\hline
   5 & 9 & 22 & 1 & 0 & 1 & 0 & 49.46 \\ 
   5 & 9 & 23 & 1 & 0 & 0 & $-$1 & 47.67 \\ 
   5 & 9 & 24 & 1 & 0 & $-$1 & 1 & 55.59 \\ 
  \cline{2-8}
	 5 & 10 & 25 & 1 & 1 & 0 & $-$1 & 43.34 \\ 
   5 & 10 & 26 & 1 & 1 & $-$1 & 0 & 48.88 \\ 
   5 & 10 & 27 & 1 & 1 & 1 & 1 & 51.32 \\ 
	\hline
   6 & 11 & 28 & $-$1 & 1 & 0 & 0 & 43.68 \\ 
   6 & 11 & 29 & $-$1 & 1 & $-$1 & 1 & 44.96 \\ 
   6 & 11 & 30 & $-$1 & 1 & 1 & $-$1 & 44.26 \\ 
  \cline{2-8}
	 6 & 12 & 14 & $-$1 & 0 & 1 & 1 & 38.21 \\ 
   6 & 12 & 13 & $-$1 & 0 & $-$1 & $-$1 & 38.72 \\ 
   6 & 12 & 15 & $-$1 & 0 & 0 & 0 & 39.02 \\ 
  \hline
\end{tabular}
\end{center}
\end{table}

\spacingset{1.45} 

In the same way as for split-plot designs, it is possible to obtain pure error estimates from other designs, such as split-split-plot designs. To illustrate this, we created artificial data for an I-optimal split-split-plot design produced by the software package JMP. The true fixed effects parameters were the same as in Example 2 and the variance components were $\sigma_1^2=4$, $\sigma_2^2=2$ and $\sigma^2=1$. The split-split-plot design involves six whole plots, twelve subplots and 36 runs. Every whole plot has two subplots of three runs. The I-optimal design, which minimizes the average variance of prediction, includes 30 distinct design points and hence 30 different treatments. The design and the treatment labels are shown in Table~\ref{sspdata}, along with the simulated data.

\begin{table}
\caption{Fixed-effect estimates and standard errors for simulated data from the design in Table~\ref{sspdata}.}
\label{tab:splitsplitres}
\begin{center}
{\footnotesize
\begin{tabular}{crr|cccc}
  \hline
\multirow{2}{*}{Parameter} & \multicolumn{2}{c}{Estimate} & \multicolumn{4}{c}{Standard error} \\
& RS-REML & PE-REML & RS-REML & PE-REML & RS-REML-KR & PE-REML-KR \\ 
  \hline
$\beta_1$ & 6.6134 & 6.6134 & 0.5340 & 0.5410 & 0.5340 & 0.5410 \\ 
  $\beta_2$ & 2.8402 & 2.8427 & 0.3856 & 0.4256 & 0.5499 & 0.6250 \\ 
  $\beta_3$ & 0.0218 & 0.0387 & 0.2310 & 0.2014 & 0.2391 & 0.2051 \\ 
  $\beta_4$ & 0.1216 & 0.1046 & 0.2310 & 0.2014 & 0.2391 & 0.2051 \\ 
  $\beta_{11}$ & $-$4.5637 & $-$4.5452 & 0.9322 & 0.9430 & 0.9362 & 0.9454 \\ 
  $\beta_{22}$ & $-$1.9252 & $-$1.8964 & 0.5460 & 0.6025 & 0.7756 & 0.8812 \\ 
  $\beta_{33}$ & 0.1064 & 0.0969 & 0.3995 & 0.3474 & 0.4023 & 0.3495 \\ 
  $\beta_{44}$ & 0.5142 & 0.5048 & 0.3932 & 0.3419 & 0.3959 & 0.3440 \\ 
  $\beta_{12}$ & $-$3.8645 & $-$3.9355 & 0.5125 & 0.5599 & 0.8285 & 0.9257 \\ 
  $\beta_{13}$ & $-$0.8496 & $-$0.8420 & 0.2742 & 0.2386 & 0.2769 & 0.2404 \\ 
  $\beta_{14}$ & 2.1437 & 2.1439 & 0.2759 & 0.2397 & 0.2760 & 0.2398 \\ 
  $\beta_{23}$ & $-$0.0526 & $-$0.0526 & 0.3107 & 0.2700 & 0.3107 & 0.2700 \\ 
  $\beta_{24}$ & 3.2443 & 3.2443 & 0.3107 & 0.2700 & 0.3107 & 0.2700 \\ 
  $\beta_{34}$ & $-$1.3678 & $-$1.4290 & 0.3152 & 0.2944 & 0.4401 & 0.3776 \\ 
   \hline
\end{tabular}}
\end{center}
\end{table}

The fixed effects' estimates obtained and their estimated standard errors are shown in Table \ref{tab:splitsplitres}. The estimates of the whole-plot, subplot and subsubplot variance components, $\sigma_1^2$, $\sigma_2^2$ and $\sigma^2$ were, respectively, 0.799, 0.296 and 1.159 from the RS-REML analysis and 0.743, 0.565 and 0.874 from the PE-REML analysis. This explains the differences we can see in the estimation of the effects of the treatment factors, though none of them is large. Again, differences in the standard errors, especially of quadratic effects, are larger, but not huge.

\section{Comparison of Methods}\label{sec:simulation}

The examples in the previous section illustrated several interesting points about the two different REML methods: in Example 1, PE-REML gave considerably smaller standard errors of some fixed effects; in Example 2, the methods gave quite different estimates of the variance components; Example 3 showed slightly different estimates of all fixed effects being obtained from the two methods. However, no general conclusions can be drawn from these examples about which method is to be preferred. To do this, we conducted a simulation study using the design from the artificial split-plot example in Section~\ref{fakeex2}. Initially, we assume that the second-order polynomial model is correct. This, of course, is unrealistic, but allows us to compare the new PE-REML method with the usual RS-REML method in the situation that is most favorable for the latter.

\subsection{Assumed Model is Correct}\label{boatymcboatface}

\begin{table}
\caption{Empirical standard errors of quadratic parameter estimates estimated from simulations when $\sigma_1^2= 4$ and $\sigma^2= 2$} \label{tab:sdests}
\begin{center}
\begin{tabular}{rrr}
  \hline
 & PE-REML & RS-REML \\
  \hline
  $\beta_{11}$ & 1.2939 & 1.2937 \\
  $\beta_{22}$ & 1.2830 & 1.2832 \\
  $\beta_{33}$ & 0.4074 & 0.4073 \\
  $\beta_{44}$ & 0.4117 & 0.4117 \\
   \hline
\end{tabular}
\end{center}
\end{table}

We simulated 10,000 data sets from the normal distribution, using the design in Table \ref{fakedata2}, assuming the same model as used in Section \ref{fakeex2}. We analyze each simulated data set using the RS-REML and PE-REML methods. The mean estimated values of $\sigma_1^2$ and $\sigma^2$ were 4.0215 and 2.0021 respectively from the RS-REML method and 4.1180 and 1.9993 from the PE-REML method. The design has the property that all subplot factor effects, except the quadratic effects, are estimated orthogonally to blocks. For this reason, the only true standard errors of fixed effects that differ between PE-REML and RS-REML are those of the quadratic effects. The empirical standard errors for these effects, calculated as the sample standard deviations of the parameter estimates in the simulations, are shown in Table \ref{tab:sdests}. In this case, where all necessary assumptions are known to be true, using the variance component estimates from the polynomial model should result in the most precise parameter estimates for the quadratic effects. The difference in precision with the PE-REML method is, however, almost nonexistent. The biases of the fixed effects estimators were also estimated from the simulations; they are similar for both methods and never more than 2.31\% of the corresponding empirical standard error. For this setup, it is clear that either method is acceptable for estimating the fixed effects parameters.

\begin{table}
\caption{Relative biases (\%) of uncorrected estimated standard errors of fixed effects when $\sigma_1^2= 4$ and $\sigma^2= 2$.} \label{tab:biasSE}
\begin{center}
\begin{tabular}{crr}
  \hline
 & PE-REML & RS-REML \\
  \hline
  $\beta_1$    & $-$7.98 & $-$3.84 \\
  $\beta_2$    & $-$7.83 & $-$3.68 \\
  $\beta_3$    & $-$2.38 &  0.19 \\
  $\beta_4$    & $-$2.78 & $-$0.22 \\
  $\beta_{11}$ & $-$8.53 & $-$4.42 \\
  $\beta_{22}$ & $-$7.75 & $-$3.64 \\
  $\beta_{33}$ & $-$8.49 & $-$1.42 \\
  $\beta_{44}$ & $-$9.73 & $-$2.47 \\
  $\beta_{12}$ & $-$7.32 & $-$3.15 \\
  $\beta_{13}$ & $-$3.15 & $-$0.60 \\
  $\beta_{14}$ & $-$3.77 & $-$1.24 \\
  $\beta_{23}$ & $-$4.01 & $-$1.48 \\
  $\beta_{24}$ & $-$3.97 & $-$1.44 \\
  $\beta_{34}$ & $-$3.89 & $-$1.35 \\
   \hline
\end{tabular}
\end{center}
\end{table}

\begin{table}
\caption{Relative biases (\%) of estimated standard errors for quadratic effects, corrected using the Kenward-Roger method, when $\sigma_1^2= 4$ and $\sigma^2= 2$.} \label{tab:biasadjSE}
\begin{center}
\begin{tabular}{rrr}
  \hline
 & PE-REML & RS-REML \\
  \hline
  $\beta_{11}$ & $-$8.51 & $-$4.42 \\
  $\beta_{22}$ & $-$7.74 & $-$3.64 \\
  $\beta_{33}$ & $-$6.37 & $-$0.70 \\
  $\beta_{44}$ & $-$7.64 & $-$1.75 \\
   \hline
\end{tabular}
\end{center}
\end{table}

It is generally more difficult to get good estimators of the standard errors of the fixed effects and these are usually biased. The biases estimated from the simulations, expressed as percentages of the corresponding empirical standard errors, are shown in Table \ref{tab:biasSE}. Because $(\mathbf{X}^\prime\hat{\bfSigma}^{-1}\mathbf{X})^{-1}$ underestimates the true variance, the biases are generally negative. They are small but non-negligible and the biases from the PE-REML method are larger than those from the RS-REML method. This is not surprising, as there are fewer residual degrees of freedom in each stratum when using the full treatment model to obtain variance component estimates, and the estimated standard errors are asymptotically unbiased as these degrees of freedom go to infinity. The Kenward-Roger correction was applied and, for the quadratic effects where it makes some difference, the results are shown in Table~\ref{tab:biasadjSE}. For both methods, the correction works well for the quadratic effects of the subplot factors ($\beta_{33}$ and $\beta_{44}$), but less well for the quadratic effects of the whole-plot factors ($\beta_{11}$ and $\beta_{22}$). 

\subsection{Assumed Model is Incorrect}

\begin{table}
\caption{Relative biases (\%) of estimated standard errors of fixed effects, corrected using the Kenward-Roger method, when the true model has a third-order term with a large effect estimated in the whole-plot stratum.} \label{tab:bias3rdorder}
\begin{center}
\begin{tabular}{crr}
  \hline
 & PE-REML & RS-REML \\
  \hline
$\beta_1$    &  $-$8.99 & 46.35 \\
$\beta_2$    &  $-$9.42 & 45.66 \\
$\beta_3$    &  $-$3.75 & $-$1.11 \\
$\beta_4$    &  $-$2.55 &  0.13 \\
$\beta_{11}$ &  $-$8.94 & 46.26 \\
$\beta_{22}$ &  $-$9.23 & 45.91 \\
$\beta_{33}$ &  $-$6.60 &  0.38 \\
$\beta_{44}$ &  $-$8.10 & $-$1.12 \\
$\beta_{12}$ &  $-$8.70 & 46.81 \\
$\beta_{13}$ &  $-$4.28 & $-$1.65 \\
$\beta_{14}$ &  $-$3.17 & $-$0.51 \\
$\beta_{23}$ &  $-$3.34 & $-$0.68 \\
$\beta_{24}$ &  $-$1.86 &  0.83 \\
$\beta_{34}$ &  $-$3.32 & $-$0.66 \\
   \hline
\end{tabular}
\end{center}
\end{table}

In Section~\ref{boatymcboatface}, the results from using the polynomial model to estimate the variance components showed smaller biases than the results from using the full treatment model. This is not surprising since the data were simulated from this polynomial model. However, if the polynomial model is wrong, the results can change drastically. In Table \ref{tab:bias3rdorder}, the relative biases of estimated standard errors are shown from additional simulations in which the quadratic by linear interaction effect $\beta_{112}$, i.e.\ the effect of $X_1^2 X_2$, was given the value $5$. The assumed third-order effect was thus similar in size to the other active effects. Also, the model misspecification involves whole-plot factors only, so that we should expect that the whole-plot aspects of the analysis are affected, rather than the subplot parts of the analysis.

It can be seen that the estimated standard errors obtained when using the pure error variance components are quite robust to this model misspecification (having a relative negative bias of no more than about $10\%$), while those based on the polynomial model fail completely for the linear effects $\beta_1$ and $\beta_2$, the quadratic effects $\beta_{11}$ and $\beta_{22}$, and the interaction effect $\beta_{12}$ (with relative positive biases greater than 40\%). This is due to the fact that the variance component $\sigma^2_1$ is overestimated substantially by the RS-REML method. The RS-REML method produced a mean estimate of 9.6323, compared with 4.0358 from the PE-REML method. This  results in inflated standard errors for the effects estimated in the whole-plot stratum. The estimates of $\sigma^2$ have means 1.9980 and 2.0091 from the PE- and RS-REML methods respectively.

The importance of the overestimated standard errors is not only that we might draw wrong conclusions about specific effects, but that we can easily be led to believe that there are few active effects and potentially miss factors which could be very important for process or product improvement. Note that a negative bias of 10\% in the estimated standard error will lead to a 5\% significance test for that parameter having a true size of 6.69\%, while a positive bias of 40\% leads to a true size of 1.77\%.

\begin{table}
\caption{Relative biases (\%) of estimated standard errors of fixed effects, corrected using the Kenward-Roger method, when the true model has a third-order term with a large effect estimated in the subplot stratum.} \label{tab:bias3rdordersubplot}
\begin{center}
\begin{tabular}{crr}
  \hline
 & PE-REML & RS-REML \\
  \hline
$\beta_1$    &  $-$9.02 & $-$4.50 \\
$\beta_2$    &  $-$9.33 & $-$4.83 \\
$\beta_3$    &  $-$3.47 & 87.66 \\
$\beta_4$    &  $-$2.65 & 89.25 \\
$\beta_{11}$ &  $-$8.59 & $-$3.85 \\
$\beta_{22}$ &  $-$8.81 & $-$4.05 \\
$\beta_{33}$ &  $-$5.97 & 76.20 \\
$\beta_{44}$ &  $-$6.00 & 75.99 \\
$\beta_{12}$ &  $-$9.12 & $-$4.61 \\
$\beta_{13}$ &  $-$2.67 & 89.22 \\
$\beta_{14}$ &  $-$4.12 & 86.40 \\
$\beta_{23}$ &  $-$3.32 & 87.95 \\
$\beta_{24}$ &  $-$3.31 & 87.97 \\
$\beta_{34}$ &  $-$4.73 & 85.21 \\
   \hline
\end{tabular}
\end{center}
\end{table}

Simulations were also run with the effect $\beta_{334}$, i.e.\ the effect of $X_3^2X_4$, which would be estimated in the subplot stratum, having a value of 5. This is a scenario in which the model misspecification is in the subplot stratum rather than in the whole-plot stratum. On average, the RS-REML method estimated $\sigma_1^2$ and $\sigma^2$ to be 2.8964 and 7.0989, whereas PE-REML gave estimates of 4.0006 and 1.9868 (close to the true values of the two variance components). The biases of the estimated standard errors, shown in Table \ref{tab:bias3rdordersubplot}, are unacceptable for all parameters estimated in the subplots stratum if the RS-REML method is used.

\begin{table}
\caption{Relative biases (\%) of estimated standard errors of fixed effects, corrected using the Kenward-Roger method, when third-order terms have small but non-zero effects.} \label{tab:biasmany3rd}
\begin{center}
\begin{tabular}{crr}
  \hline
 & PE-REML & RS-REML \\
  \hline
$\beta_1$    & $-$7.36 &  1.59 \\
$\beta_2$    & $-$8.97 & $-$0.17 \\
$\beta_3$    & $-$2.87 & 19.77 \\
$\beta_4$    & $-$3.03 & 19.58 \\
$\beta_{11}$ & $-$8.80 &  0.03 \\
$\beta_{22}$ & $-$8.52 &  0.32 \\
$\beta_{33}$ & $-$5.79 & 19.83 \\
$\beta_{44}$ & $-$7.06 & 18.18 \\
$\beta_{12}$ & $-$8.39 &  0.47 \\
$\beta_{13}$ & $-$4.15 & 18.19 \\
$\beta_{14}$ & $-$2.57 & 20.14 \\
$\beta_{23}$ & $-$2.79 & 19.87 \\
$\beta_{24}$ & $-$3.13 & 19.46 \\
$\beta_{34}$ & $-$3.26 & 19.29 \\
   \hline
\end{tabular}
\end{center}
\end{table}

Some simulations were also run in which all third-order terms except the pure cubic terms had small, but non-zero, effects of size $0.5$ for linear by quadratic interactions and $0.25$ for linear by linear by linear interactions. The result was that $\sigma_1^2$ and $\sigma^2$ were estimated to be 4.0483 and 1.9888 respectively by PE-REML and 4.2139 and 2.8793 by RS-REML. The biases of the estimated standard errors of the fixed effects, shown in Table \ref{tab:biasmany3rd}, indicate that, in general, it is clearly better to use the pure error estimates of the variance components. Using that approach, the relative biases are negative and less than 10\%, compared with positive biases of up to 20\% when using the RS-REML method. This time, it is mainly the estimate for $\sigma^2$ which is inflated, resulting in substantial upward biases for the estimated standard errors of effects estimated in the subplot stratum when using the RS-REML method. A positive bias of 20\% for the standard errors corresponds to a size of 2.91\% for a significance test at the 5\% level.

\section{Discussion}\label{sec:discussion}

On the basis of our results, we strongly recommend that the variance components in multi-stratum response surface designs should be routinely estimated using PE-REML, and thus based on the full treatment model. REML is implemented in many statistical packages and generally has good properties. Implementing PE-REML simply involves running the available REML procedure with the full set of treatment indicators as the fixed effects, to obtain the estimates of the variance components. These are then plugged in to the generalized least squares formula to obtain the estimates of the fixed effects parameters. The same method could be used for analyzing data from blocked experiments, though in most cases it will make little difference, since most information on treatment effects comes from within the blocks.

The results in Section \ref{sec:simulation} show that PE-REML gives stable estimates of the fixed effects parameters and their standard errors, irrespective of whether or not the assumed model is correct. The standard errors are consistently negatively biased, by up to about 10\% in relative terms. This is not a major concern, but it does mean that aspects of inference, such as p-values and interval estimates at a given level of confidence are not exact. In this case, the Kenward-Roger correction helps only a little and looking for better correction methods might be a fruitful avenue for further research. 

The RS-REML method is highly sensitive to misspecification of the polynomial response surface model. The standard errors resulting from that method are biased upward substantially in the event of misspecification of the model, rendering the detection of active effects much more difficult. Therefore, we recommend the PE-REML method presented here, whenever the design of the experiment allows pure error estimates of the variance components.

\section*{Appendix}

The derivation of results needed to calculate the Kenward-Roger correction for PE-REML in the split-split-plot design follows the same steps as for the split-plot design, given in Section \ref{sec:SEs}, though, of course, there is an additional variance component. In a split-split-plot design with $b$ subplots within each whole plot and $k$ subsubplots within each subplot, $\bfsigma^\prime = [\sigma_1^2~\sigma_2^2~\sigma^2]$ and
\[
\bfSigma = \sigma^2\mathbf{I} +\sigma_1^2\mathbf{Z}_1\mathbf{Z}_1^\prime +\sigma_2^2\mathbf{Z}_2\mathbf{Z}_2^\prime.
\]
By twice applying the formula for Schur complements, as used in the derivation of (\ref{eq:SigmaInv}), we obtain
\[
\bfSigma^{-1} = \frac{1}{\sigma^2}\left\{\mathbf{I} -{\sigma^2\sigma_1^2 \over (\sigma^2+\sigma_2^2k)(\sigma^2+\sigma_1^2bk+\sigma_2^2k)}\mathbf{Z}_1\mathbf{Z}_1^\prime -{\sigma_2^2 \over \sigma^2+\sigma_2^2k}\mathbf{Z}_2\mathbf{Z}_2^\prime\right\}.
\]
Differentiating with respect to each variance component and simplifying, we obtain
\[
{\partial\bfSigma^{-1} \over \partial\sigma_1^2} = -{1 \over (\sigma^2+\sigma_1^2bk+\sigma_2^2k)^2}\mathbf{Z}_1\mathbf{Z}_1^\prime,
\]
\[
{\partial\bfSigma^{-1} \over \partial\sigma_2^2} = {1 \over (\sigma^2+\sigma_2^2k)^2}\left\{{\sigma_1^2k(2\sigma^2+2k\sigma_2^2kb\sigma_1^2) \over (\sigma^2+\sigma_1^2bk+\sigma_2^2k)^2}\mathbf{Z}_1\mathbf{Z}_1^\prime -\mathbf{Z}_2\mathbf{Z}_2^\prime\right\}
\]
and
\[
{\partial\bfSigma^{-1} \over \partial\sigma^2} = \frac{1}{\sigma^2}\left\{{\sigma_1^2(2\sigma^4+2\sigma^2\sigma_2^2k+\sigma_1^2\sigma_2^2bk^2) \over (\sigma^2+\sigma_2^2k)^2(\sigma^2+\sigma_1^2bk+\sigma_2^2k)^2}\mathbf{Z}_1\mathbf{Z}_1^\prime +{\sigma_2^2(2\sigma^2+\sigma_2^2k) \over \sigma^2(\sigma^2+\sigma_2^2k)^2}\mathbf{Z}_2\mathbf{Z}_2^\prime -\frac{1}{\sigma^2}\mathbf{I}\right\}.
\]
The elements of the matrix $\mathbf{U}$ are obtained by direct numerical inversion of the diagonal block of the Fisher information matrix corresponding to the variance components - see \citet[p.\ 176-178]{mcsene}.

\section*{Acknowledgements}

This work was supported by the Royal Society International Joint Project number 2007/R2 and EPSRC grant number EP/T021624/1. Part of it was completed while the authors were participants in the Isaac Newton Institute for Mathematical Sciences program on the Design and Analysis of Experiments.

\bibliographystyle{chicago}
\bibliography{newREML3}

\end{document}